\documentclass[journal=jacsat,manuscript=article]{achemso}

\usepackage{graphicx}
\usepackage{color}
\usepackage{amsmath}

\newcommand{\kmos}{K$_{0.4}$MoS$_2$}
\newcommand{\mos}{MoS$_2$}
\newcommand{\nbse}{NbSe$_2$}
\newcommand{\obo}{$1\times1$}
\newcommand{\tbt}{$2\times2$}
\newcommand{\thbth}{$3\times3$}
\newcommand{\trt}{$2\sqrt{3}$}%{$2\sqrt{3}\times2\sqrt{3}$}

\newcommand{\tc}{T_\text{c}}
\newcommand{\tcdw}{T_\text{CDW}}
\newcommand{\ntd}{n_\text{2D}}

\newcommand{\hl}[1]{#1}

\author{Mohammed K. Bin Subhan}
\affiliation{Department of Physics and Astronomy, University College London, WC1E 6BT, London, UK}

\author{Asif Suleman}
\affiliation{Department of Physics and Astronomy, University College London, WC1E 6BT, London, UK}
\altaffiliation{London Centre for Nanotechnology, University College London, WC1H 0AH, London, UK}

\author{Gareth Moore}
\affiliation{Department of Physics and Astronomy, University College London, WC1E 6BT, London, UK}
\altaffiliation{London Centre for Nanotechnology, University College London, WC1H 0AH, London, UK}

\author{Peter Phu}
\affiliation{Department of Physics and Astronomy, University College London, WC1E 6BT, London, UK}

\author{Moritz Hoesch}
\affiliation{Photon Science, Deutsches Elektronen-Synchrotron (DESY), Notkestrasse 85, 22607 Hamburg, Germany}
\author{Hidekazu Kurebayashi}
\affiliation{London Centre for Nanotechnology, University College London, WC1H 0AH, London, UK}
\altaffiliation{Department of Electronic and Electrical Engineering, University College London, WC1E 6BT, London, UK}

\author{Christopher A. Howard}
\affiliation{Department of Physics and Astronomy, University College London, WC1E 6BT, London, UK}
\email{c.howard@ucl.ac.uk}

\author{Steven~R.~Schofield}
\affiliation{Department of Physics and Astronomy, University College London, WC1E 6BT, London, UK}
\altaffiliation{London Centre for Nanotechnology, University College London, WC1H 0AH, London, UK}
\email{s.schofield@ucl.ac.uk}

\title{Charge density waves in electron-doped molybdenum disulfide}


\begin{document}

\date{\today}

\begin{abstract}
We present the discovery of a charge density wave (CDW) ground state in heavily electron-doped molybdenum disulfide (\mos). This is the first observation of a CDW in any $d^2$ (column 6) transition metal dichalcogenide (TMD). The band structure of \mos\ is distinct from the $d^0$ and $d^1$ TMDs in which CDWs have been previously observed, facilitating new insight into CDW formation.  We demonstrate a metal-insulator transition at 85~K, a 25~meV gap at the Fermi level, and two distinct CDW modulations, $(2\sqrt{3}\times2\sqrt{3})$R$30^\circ$ and \tbt, attributable to Fermi surface nesting (FSN) and electron-phonon coupling (EPC), respectively. This simultaneous exhibition of FSN and EPC CDW modulations is unique among observations of CDW ground states, and we discuss this in the context of band folding.  Our observations provide a route toward the resolution of controversies surrounding the origin of CDW modulations in TMDs.
\end{abstract}

Strongly anisotropic crystals that confine charge carriers to two-dimensions (2D) exhibit a rich diversity of correlated ground states, including charge density waves (CDWs), spin density waves, and superconductivity. However, despite decades of intense effort, there are still large gaps in our understanding of the mechanisms underpinning the formation and competition between such states. New experimental observations of correlated states can provide litmus tests for competing theoretical models. CDWs are a periodic spatial oscillation of charge density, accompanied by a lattice distortion, that occur in crystalline materials due to electron-electron and electron-phonon interactions~\cite{Gruner1994,Zhu2015,Rossnagel2011}.  Despite intense investigation, the physics of CDW formation remains a topic of vigorous debate~\cite{Ugeda2016,Zhu2017}, and the connection to other exotic electronic ground states, most notably superconductivity, remains controversial~\cite{Gabovich2002}.  

Transition metal dichalcogenides (TMDs) are two-dimensional (2D) layered materials that are tailorable by varying the elemental composition, coordination, symmetry, layer number and separation, and doping~\cite{Chhowalla2013,Manzeli2017}. TMDs thus provide excellent opportunities to investigate fundamental condensed matter physics in reduced dimensions. CDWs have been discovered in the semimetallic column 4 ($d^0$) TMDs, TiSe$_2$~\cite{Sugawara2015}, and TiTe$_2$~\cite{Chen2017}, and in the metallic column 5 ($d^1$) TMDs VS$_2$~\cite{Mulazzi2010}, VSe$_2$~\cite{Strocov2012}, NbSe$_2$~\cite{Ugeda2016,Arguello2014}, TaS$_2$~\cite{Wilson2001}, and TaSe$_2$\cite{Liu2000}.  CDWs have not been previously observed in column 6 ($d^2$) TMDs, which are typically band semiconductors. However, density functional theory calculations have indicted the possibility of CDW formation in \hl{heavily-doped bulk~\cite{Enyashin2012, Chen2013} and monolayer \mos,~\cite{Rosner2014,Zhuang2017}} and a recent study reported anomalies in the temperature dependence of the sheet resistance in electron-doped \mos, suggesting the possibility of a CDW phase transition~\cite{Piatti2018}. There are two popular mechanisms for CDW formation in TMDs: Fermi surface nesting (FSN; the favoured model in VSe$_2$, TaS$_2$, and TaSe$_2$~\cite{Rossnagel2011,Strocov2012}), which requires coupling of the Fermi surface and leads to the opening of a small energy gap centred at the Fermi energy; and momentum-dependent electron-phonon coupling (EPC; favoured in NbSe$_2$ and VS$_2$~\cite{Mulazzi2010}), which can occur in the absence of strong connections within the Fermi surface. In many cases, not all aspects of the data are adequately described by either model.  

\begin{figure}[t!]
\centering
\includegraphics[width=12cm]{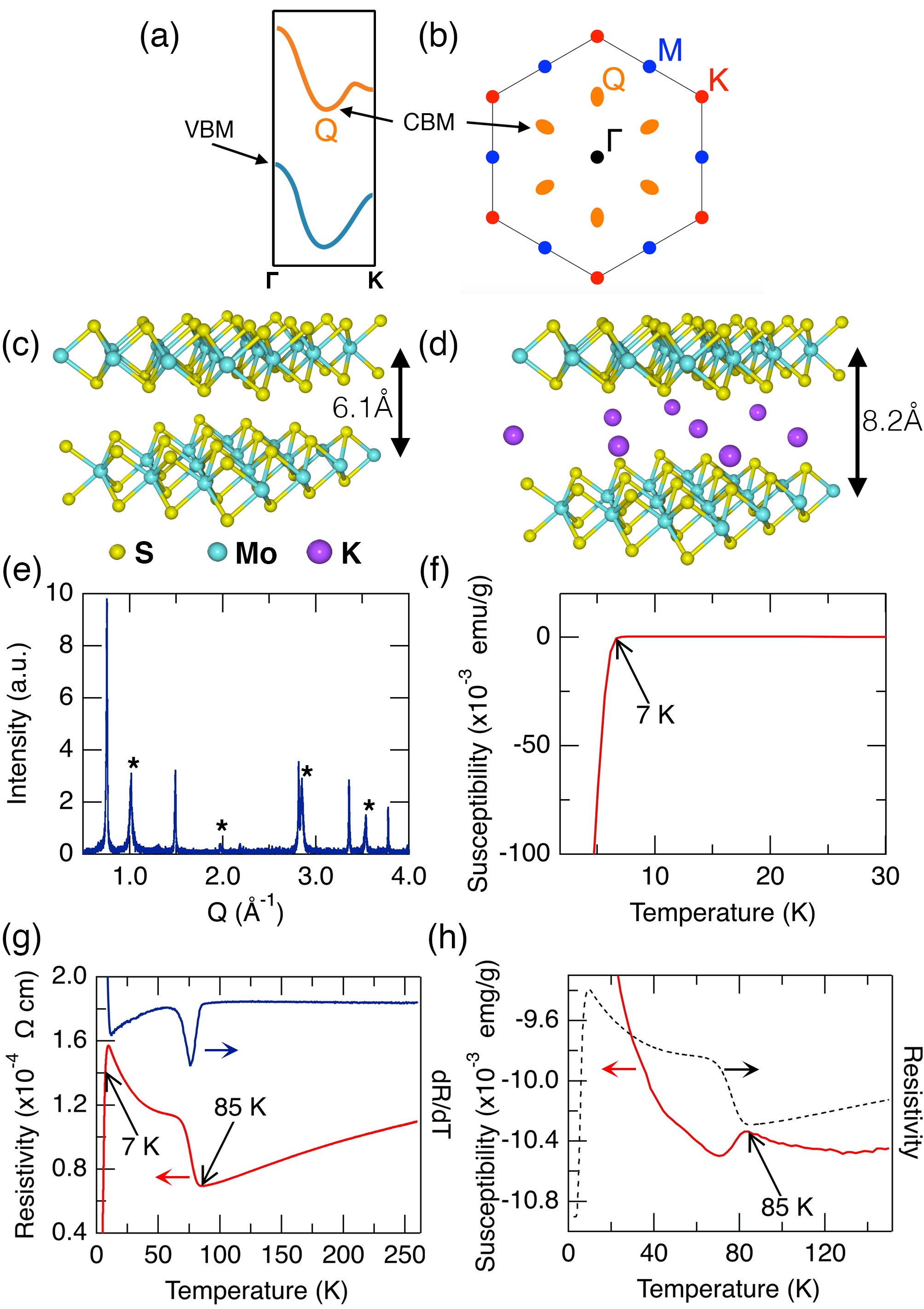}
\caption{(a) \mos\ band structure schematic highlighting valence/conduction band maximum/minimum (VBM/CBM) (adapted from Ref.~\cite{Splendiani2010}).  (b) Brillouin zone with high-symmmetry points $\Gamma$, K, M, and low-symmetry point Q.  (c) Crystal structure of \mos\ and (d) of \kmos.  (e) \kmos\ x-ray diffraction; asterisks denote unintercalated \mos\ peaks.  (f) Magnetic susceptibility (10~Oersted). (g) \hl{Resistivity} (red trace) showing metal-insulator ($85$~K) and superconducting transitions ($7$~K); $dR_\text{S}/dT$ is shown on the right axis (blue trace). (h) Magnetic susceptibility (red trace) with a large applied field ($10^4$ Oe) with \hl{resistivity} data overlayed (dashed curve; right axis).} 
\label{fig1}
\end{figure}

\mos\ is a band semiconductor, column 6 ($d^2$) TMD with a trigonal prismatic (2H) ground state (Fig.~\ref{fig1}c)~\cite{Chhowalla2013}. The band edges derive from $4d$ orbitals~\cite{Chhowalla2013}, and the bulk material has a 1.29~eV indirect band gap~\cite{Mak2010} (Fig.~\ref{fig1}a).  In reciprocal space, the valence band maximum is at the zone centre ($\Gamma$), while the conduction band minimum is located midway along $\Gamma$--K, producing a sixfold degenerate conduction band with electron pockets at the points labelled Q in Figs.~\ref{fig1}a,b. With electron doping, \mos\ undergoes a metal-insulator transition at a free carrier density of $\ntd = 6.7 \times 10^{12}$~cm$^{-2}$, and exhibits a sharp onset of superconductivity at $6.8\times10^{13}$~cm$^{-2}$~\cite{Ye2012}.  

Here, we present the discovery of a CDW ground state in bulk potassium-intercalated (and therefore electron-doped) \mos\ with a simultaneous exhibition of FSN and EPC derived modulations. We demonstrate a metal-insulator transition at 85~K, $(2\sqrt{3}\times2\sqrt{3})$R$30^\circ$ and $2\times2$ CDW modulations via atomic-resolution scanning tunnelling microscopy (STM), and a 25~meV energy gap at the Fermi level via tunnelling spectroscopy. The $2\sqrt{3}$ modulation is perfectly matched by a nesting vector connecting the conduction band pockets, while the $2\times2$ modulation matches a theoretically predicted phonon-mode softening at the M point~\cite{Rosner2014}. We discuss that the two modulations are simple linear combinations of one another, suggesting that the driving mechanisms may be coupled via band folding.  

Potassium ions were intercalated into the Van der Waals gaps of a bulk \mos\ sample using the well-established low-temperature liquid ammonia method~\cite{Zhang2016}.  Briefly, high-quality \mos\ crystals (Manchester Nanomaterials) were degased (523~K, $<10^{-6}$~mbar) then combined with potassium dissolved in liquid ammonia at 218~K. Potassium intercalation (Fig.~\ref{fig1}d) completed after $\sim24$~h. X-ray diffraction (XRD) was measured in a reflection geometry (Philips X'Pert) on cleaved samples in an airtight beryllium dome.  Magnetic susceptibility measurements (Quantum Design MPMS-7) were performed on samples held in a plastic capsule and sample straw.  Four-terminal contacts were attached to the sample using Epotek H21D silver epoxy and  transport measurements (Keithley 2400 SMU and Stanford Research Systems SR830) were made on a cold finger below $10^{-5}$~mbar with a 1~mA current and 10~Kh$^{-1}$ heating rate. STM measurements (Omicron LT-STM) were performed on samples cleaved under ultrahigh vacuum ($<5\times10^{-10}$~mbar) at room temperature to produce an atomically-clean surface of a bulk intercalated sample, and then cooled to 5.5~K for STM measurement.

XRD confirms the crystalline quality of our samples (Fig.~\ref{fig1}e); the $00l$ out-of-plane peaks are shifted with respect to their unintercalated positions, demonstrating the expected $2.2\pm0.1$~\AA\ increase in the layer separation~\cite{Zhang2016}.  Magnetic susceptibility and four-terminal \hl{resistivity} measurements (Figs.~\ref{fig1}f and \ref{fig1}g) confirm the onset of superconductivity at $\tc=7.0\pm0.5$~K~\cite{Woollam1976,Zhang2016}. The room temperature \hl{resistivity ($R_\text{s}\sim1\times10^{-4}$~$\Omega$cm)} decreases linearly with temperature (Fig.~\ref{fig1}g, red trace), as expected~\cite{Lu2015}. At 85~K we find a pronounced step increase of \hl{$\sim5\times10^{-5}$~$\Omega$cm}, marking the location of a metal-insulator transition. Also at 85~K we find an abrupt decrease ($-0.02$~emu/g) in magnetic susceptibility (Fig.~\ref{fig1}h).  These features are characteristic of the opening of an energy gap at the Fermi level, and similar behaviour has been attributed to CDW transitions in VS$_2$~\cite{Mulazzi2010} and TaS$_2$~\cite{Kratochvilova2017}.

\begin{figure}[t]
\centering
\includegraphics[width=10cm]{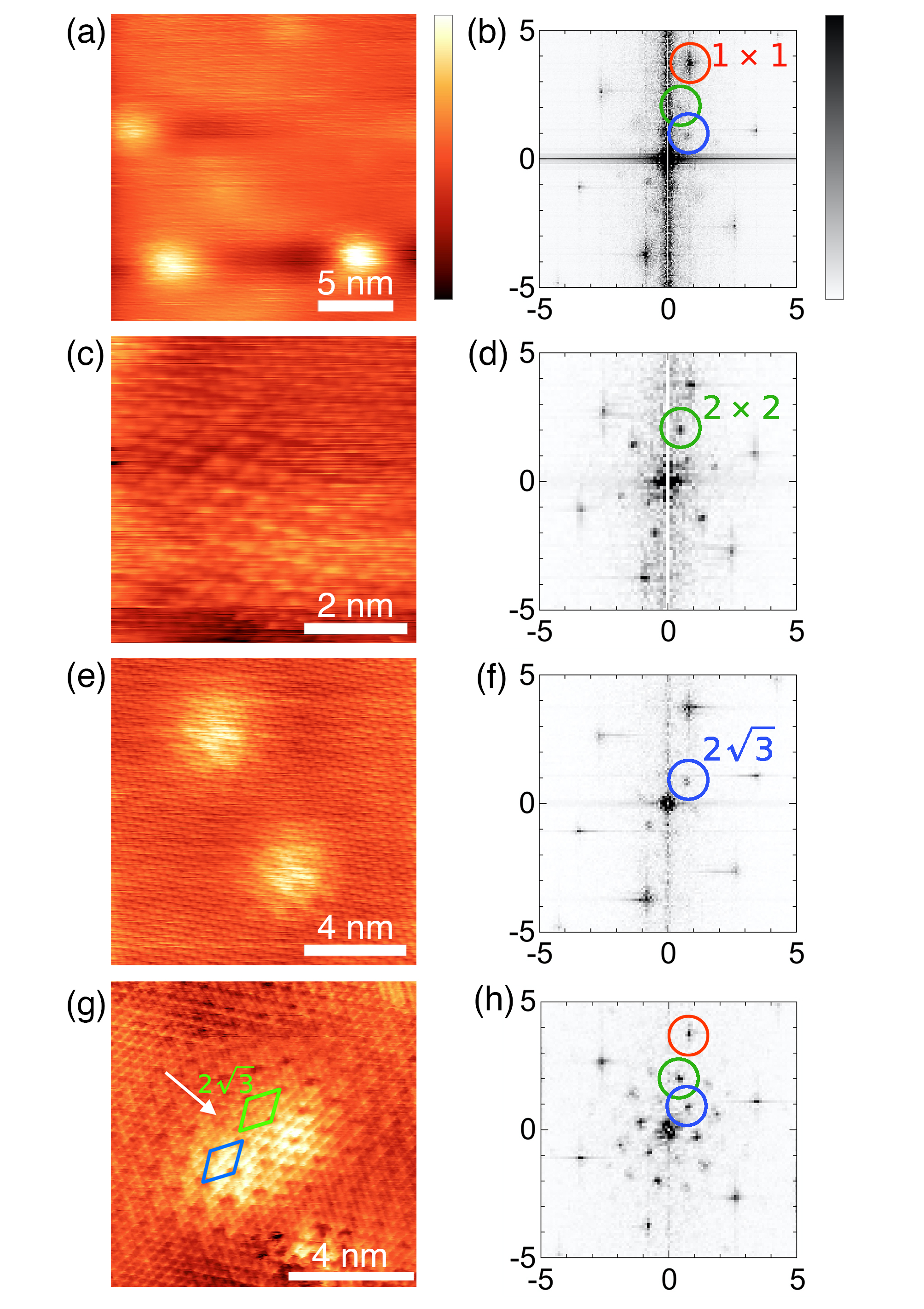}
\caption{(a,c,e)  STM topographic images of \kmos\ (5.5~K), and (b,e,f) their corresponding 2D Fourier transforms. Bragg spots corresponding to the $1\times1$ surface sulphur lattice are seen in all images, while to a lesser or greater extent \tbt\ and \trt\ periodicities are also observed (see text).  Image parameters: (a) $-150$~mV, 20~pA, z-range 1.5~nm; (c,e) $-100$~mV, 30~pA, z-range 700~pm.  FFT axes are in nm$^{-1}$ units. (g) Filled-state STM image of a pair of closely-spaced defects in \kmos.  \obo, \tbt, and \trt\ periodicities can be seen in the image and its Fourier transform in panel (h).  A phase slip boundary (white arrow) exists between the two regions of \trt\ periodicity, which we highlight on panel (g) by adding a blue rhombus to indicate the phase of the \trt\ modulation at the bottom left defect site, and a green rhombus to indicate the \trt\ phase of the top right defect.}
\label{fig:STMcdw}
\end{figure}

Figure~\ref{fig:STMcdw}a displays a topographic STM image of a region free from step edges or adatoms, and that is characteristic of images acquired using different samples and tips. The Fourier transform of this image (Fig.~\ref{fig:STMcdw}b) exhibits hexagonal spots corresponding to a lattice constant, $a=3.14\pm0.07$~\AA, in agreement with the calculated lattice constant, 3.176~\AA~\cite{Andersen2012}, confirming that we are imaging the surface sulfur atoms of the cleaved sample. A 0.2 monolayer surface coverage of potassium ions might be expected if half of the potassium ions in the cleaved layer remain on the surface after cleaving. However, we do not observe any surface potassium ions in our images and attribute this to the surface potassium ions diffusing to step edges or elsewhere while the sample is at room temperature before it is loaded into the STM and cooled. Several defects in Fig.~\ref{fig:STMcdw}a appear as protrusions several nanometres in diameter superimposed on the surface atomic lattice.  Similar defects, attributed to molybdenum vacancies or antisites, have been shown to locally enhance the \mos\ interlayer coupling~\cite{Sengoku1995,Bampoulis2017}

Figure~\ref{fig:STMcdw}c shows a higher resolution image acquired with a lower imaging bias magnitude ($-100$~mV). This image, and the corresponding Fourier transform (Fig.~\ref{fig:STMcdw}d), reveals a longer-ranged periodicity of $5.8\pm0.2$~\AA; i.e., $2\pm0.2$ times the \obo\ lattice, which we interpret as a \tbt\ CDW modulation.  We notice that the apparent intensity of the \tbt\ modulation varies within the image (Fig.~\ref{fig:STMcdw}c), suggesting a nearly-commensurate CDW phase, as observed in STM images of NbSe$_2$~\cite{Arguello2014}, and copper-intercalated TiSe$_2$~\cite{Yan2017}. Examining again the Fourier transform of the larger area image in Fig.~\ref{fig:STMcdw}b, we see the \tbt\ spots are also faintly visible in this data.  Indeed, in all images of the \kmos\ surface acquired with sufficient resolution we find evidence for a \tbt\ periodicity to a greater or lesser extent depending on the imaging parameters and location on the surface, suggesting that in \kmos\ we have a nearly-commensurate \tbt\ CDW state present everywhere throughout the sample surface.  

The image in Fig.~\ref{fig:STMcdw}e shows an area where two defects are present. Here, we find an additional periodicity, distinct from both the \obo\ lattice, and the \tbt\ modulations. The Fourier transform (Fig.~\ref{fig:STMcdw}f) demonstrates that this periodicity is characterised by a $10.4 \pm 0.2$~\AA\ real space vector, which is rotated 30$^\circ$ with respect to the \obo\ lattice.  We note that within uncertainties, this periodicity is $2\sqrt{3}$ times the \obo\ unit cell vector ($=10.7\pm0.4$), and we therefore label this as a $(2\sqrt{3}\times2\sqrt{3})\text{R}30^\circ$ modulation of the \kmos\ surface.  Unlike the \tbt\ modulation, the \trt\ modulation is strongly enhanced within the vicinity of the defect sites. We found such \trt\ modulations always occur at such defects sites; e.g., the \trt\ spots are visible in Fig~\ref{fig:STMcdw}b due to the defects in the corresponding STM image, and further examples are shown in Figs.~\ref{fig:STMcdw}g,h and Supplementary Figure S1).  Modulations of the local density of states in the vicinity of defects can occur due to quasiparticle interference (QPI)~\cite{Chen2017a}; however, we rule out QPI in the present case since the modulation is dispersionless~\cite{Chen2017a} (see the spatially-resolved tunnelling spectroscopy presented below). It is possible also to draw comparison to NbSe$_2$, where the \thbth\ CDW modulation was found to persist above $\tcdw$ in patches surrounding defects~\cite{Arguello2014}.  However, our measurements of defects in \kmos\ at 5.5, 10, and 77~K show negligible change in the spatial extent of the \trt\ modulation surrounding the defects with temperature (see Supplementary Figure S1).  Moreover, we also observe phase slip boundaries in the \trt\ modulation for closely spaced defects, similar to observations of phase boundaries in CDW modulations seen in \nbse~\cite{Soumyanarayanan2013} and TiSe$_2$~\cite{Novello2017}.  
We show an example of this in Fig.~\ref{fig:STMcdw}g, which shows a filled-state STM image where two defects are present (the corresponding Fourier transform is shown in Fig.~\ref{fig:STMcdw}h). The two defects exhibit \trt\ modulations, but these two modulations are not in phase, leading to a phase slip boundary between them, which we highlight with a white arrow in Fig.~\ref{fig:STMcdw}g. Further analysis and discussion of this phase slip boundary can be found in Supplementary Figure S2.
The implication of these observations is that the \trt\ modulation can be attributed to a CDW  whose enhancement at defect sites is an intrinsic property, and not due to QPI or an incompletely formed CDW phase.  

Other non-CDW explanations for the observed \tbt\ and \trt\ periodicities can be ruled out. The precise arrangement of the potassium ions in \kmos\ is not known~\cite{Zhang2016,Andersen2012}; however, the \kmos\ stoichiometry~\cite{Zhang2016} and hexagonal symmetry of the crystal suggests a hexagonal arrangement of the average positions of the potassium ions, with a lattice constant of $\sqrt{5/2}a$.  Thus, even allowing for all possible rotational orientations of the potassium ion layer with respect to the top \mos\ layer, we can rule out both a direct influence of the potassium ion positions, and Moir\'e interference between the \mos\ and potassium ion layers as explanations for either periodicity.  Moir\'e effects due to strain at defect sites can be additionally ruled out by the observation that the characteristic length of the modulation does not vary with distance from the defect centre as would be expected for a point-source lattice distortion derived Moir\'e interference.

\begin{figure}[t!]
\centering
\includegraphics[width=12cm]{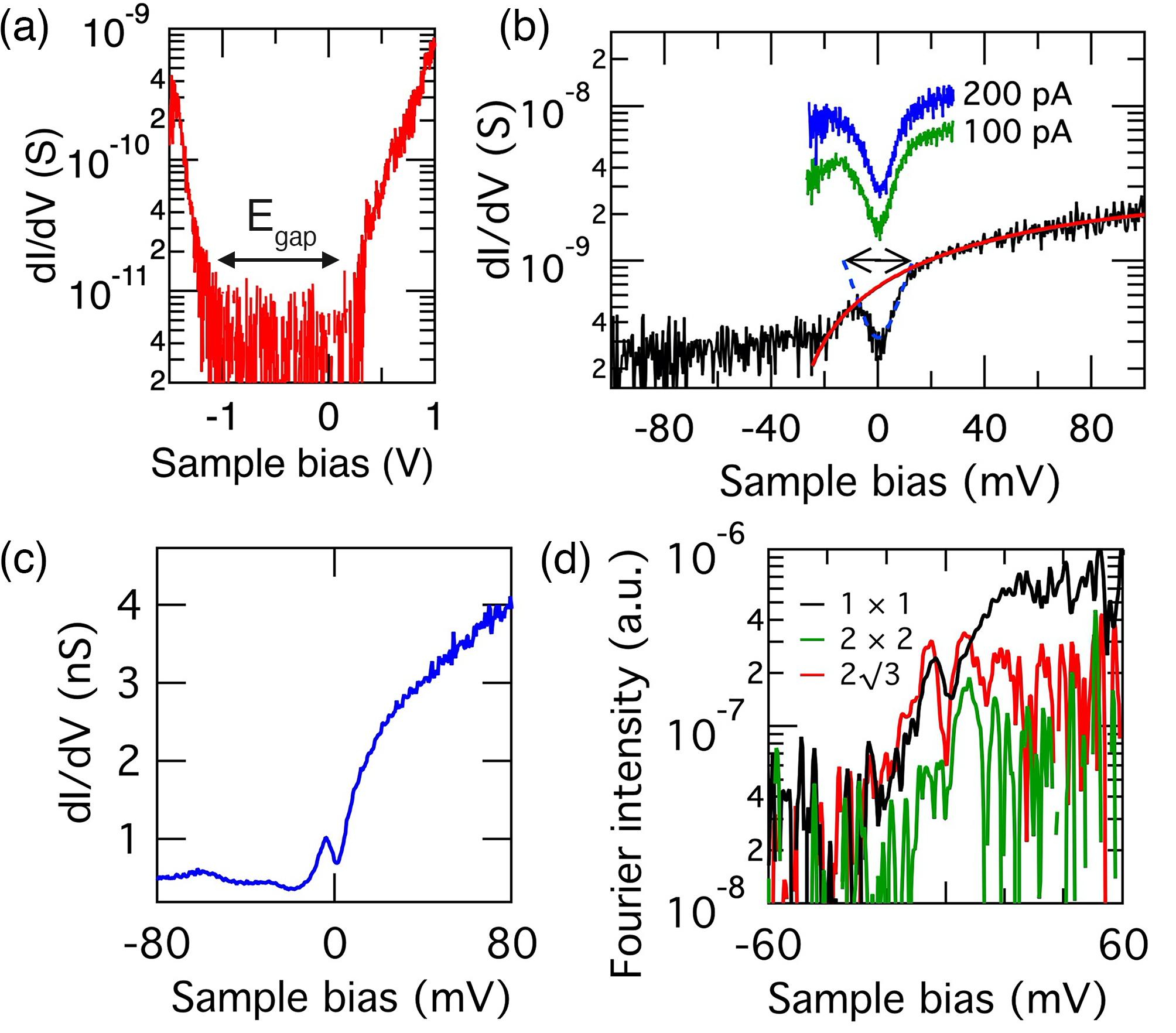}
\caption{Point tunnelling spectroscopy of \kmos\ highlighting (a) the $\sim1.2$~eV band gap, and (b) the conduction band minimum at $-25$~meV and gap at the Fermi level; curve fits are to the band edge (red) and energy gap (dashed blue).  (c) Spatial average of dI/dV.  (d) Fourier spot intensities for the \obo, \tbt, and \trt\ lattice spots (see Supplementary Figure S3 for corresponding dI/dV images at $\pm5$~mV). Tunnelling spectra were acquired by measuring current as a function of voltage with the feedback loop off. Set-point parameters: (a) $-1.5$~V, 50~pA; (b) black: $-100$~meV, 30 pA; green $-30$~meV, 100 pA; blue: $-30$~meV, 200 pA. All data 5.5~k, except panel (b) 10~K.  }
\label{fig:sts}
\end{figure}

Figure~\ref{fig:sts} presents scanning tunnelling spectroscopy (STS) measurements of \kmos. Point spectra show a bulk band gap of around 1.2~eV (Fig.~\ref{fig:sts}a), consistent with previous measurements~\cite{Bampoulis2017}, and calculations~\cite{Andersen2012}.  Higher-resolution measurements at the conduction band edge (Fig.~\ref{fig:sts}b) show the band minimum is located $\sim25$~meV below the Fermi level, in agreement with ARPES measurements of potassium-doped \mos~\cite{Eknapakul2014}, and a similar band shift in copper intercalated TiSe$_2$~\cite{Yan2017}. 

We also find an energy gap at the Fermi level \hl{(Figs.~\ref{fig:sts}b,\ref{fig:sts}c)}; we measure the the gap as the point where a quadratic fit to the gap intersects the conduction band, and take the gap to be twice this value, yielding $2\Delta=24\pm4$~meV.  There was no measurable difference between spectra taken on or away-from defect sites, demonstrating that the gap occurs across the entire surface.   

We have measured spatially-resolved conductivity, $g(r, V)$, where $r$ is the tip position and $V$ the sample bias (see Supplementary Figure S3); when integrated over $r$ (Fig.~\ref{fig:sts}c) this yields good agreement with the point spectroscopy in Fig.~\ref{fig:sts}b.  We deconvolve contributions associated with the different superlattice modulations by taking a 2D Fourier transform of $g(r, V)$ for each value of bias, to obtain conductivity as a function of the reciprocal space vector $q$. We identify in this data the expected \obo, \tbt, and \trt\ reciprocal lattice vectors, and show the intensity variation of these spots as a function of bias in Fig.~\ref{fig:sts}d.  The observation that these spots occur at the same values of $q$ for each value of bias \hl{(See Supplementary Figure S3)} rules out QPI origins for these modulations, and we do not observe any additional spots for any values of the applied bias. The \obo\ spot intensity variation (Fig.~\ref{fig:sts}d) closely matches the spatially-averaged conductance (Fig.~\ref{fig:sts}c).  
\hl{In contrast, the \trt\ spots exhibit a gap feature and peaked intensities either side of the Fermi level. The intensity variation of the \tbt\ spots follow similarly to the \trt\ spots, although the signal-to-noise ratio is insufficient to say conclusively whether or not the gap exists in the \tbt\ data.}  Concentrating on the \trt\ curve, which has the better signal-to-noise ratio, we measure peak maxima at $\pm 3.1 \pm 0.1$~mV, i.e., separated about the Fermi level by a width of $6.2\pm0.2$~mV.  Alternatively, if we measure the width from the outer edges of the peaks, we find a width of $26\pm2$~mV.  These widths compare well to the gap width estimation from our point spectra in Fig.~\ref{fig:sts}b.  Thus, the spatially-resolved tunnelling spectroscopy data shown in Fig.~\ref{fig:sts}\hl{, and the fact that the Fermi level gap in STS spectra occurs far from defect sites where the \tbt\ modulation persists but the \trt\ modulation doesn't,} suggests that the energy gap at the Fermi level correlates to both the \tbt\ and \trt\ spatial modulations observed in our STM and STS images.

\begin{figure}[t!]
\centering
\includegraphics[width=12cm]{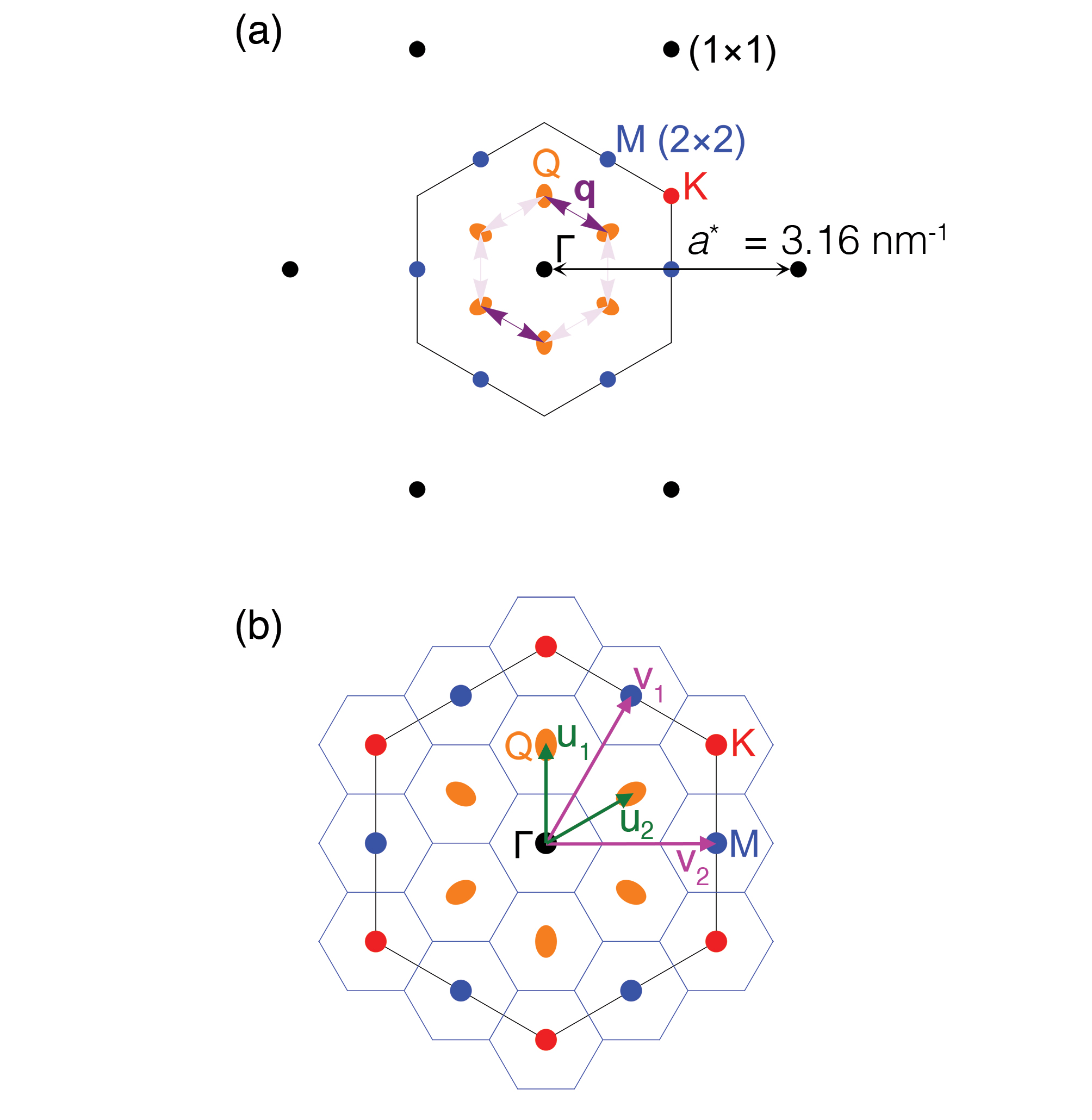}
\caption{(a) Reciprocal lattice (black spots) and Brillouin zone with high symmetry points $\Gamma$, K, and M.  Band minima at the low symmetry points, Q, lying half way along $\Gamma - \text{K}$, are shown.  (b) Brillouin zone drawn with an overlaid $(2\sqrt{3}\times2\sqrt{3})$R$30^\circ$ unit cell. Vectors $u_1, u_2$ and $v_1, v_2$ indicate the Q and M points, respectively.}
\label{fig:brill}
\end{figure}

Our observations of a metal-insulator transition at 85~K and a $24\pm4$~meV energy gap centred at the Fermi level are suggestive of a FSN-driven CDW in \kmos. The necessary FSN vectors exist in the \kmos\ electronic structure due to the fact that potassium intercalation of MoS$_2$ creates electron pockets at the Q points. The length of the  vector, $\mathbf{q}$ (Fig.~\ref{fig:brill}a), that connects the Fermi surfaces at Q is $|\boldsymbol{\Gamma} \textbf{K}|/2$, i.e., $|\mathbf{q}|= \frac{|\textbf{b}|}{4}/\cos\frac{\pi}{6}=b/2\sqrt{3}$, where $\mathbf{b}$ is the reciprocal lattice vector. There are three pairs of such nesting vectors, each rotated 30$^\circ$ to the \obo\ surface sulfur lattice. Thus, the vectors $\mathbf{q}$ connecting neighbouring Q points provide an excellent match to our observed $(2\sqrt{3}\times2\sqrt{3})\text{R}30^\circ$ modulation in STM and STS data.  This provides strong evidence for a FSN-driven \trt\ CDW phase in \kmos\ that is localised at, or enhanced by defects in the \mos\ sheets. The reason for this enhancement at defects is that the defects locally enhance the interlayer coupling,~\cite{Sengoku1995,Bampoulis2017} resulting in a lowering of the band edge at the Q point and thus an increased occupancy of the bulk-like electron pockets~\cite{Chhowalla2013} and a corresponding enhancement of the \trt\ FSN. 

However, FSN cannot explain the appearance of the \tbt\ periodicity in our STM data, because the electronic bands at the M-point are much higher in energy than the band minima~\cite{Chhowalla2013}.  However, calculations of the \mos\ phonon band structure as a function of electron doping have demonstrated phonon softening at the M point, becoming imaginary for electron doping levels exceeding $\sim0.14$ electrons per molybdenum atom.~\cite{Rosner2014,Garcia-Goiricelaya2020}   The M point lies on the Brillouin zone boundary along the reciprocal lattice vector direction (Fig.~\ref{fig:brill}a), and as such is inherently associated with a \tbt\ periodicity. This strongly suggests that momentum-dependent EPC can account for the formation of the \tbt\ phase in \kmos.

Thus, our observations of a metal-insulator transition at 85~K, the opening of a $\sim25$~meV energy gap at the Fermi level, and the real space observation of a \trt\ modulation in STM and STS data provide compelling evidence for the existence of a FSN-driven CDW phase, where the nesting vector connects electron pockets at the Q points.  This \trt\ phase is locally enhanced by defects in the \mos\ sheets that locally alter the interlayer coupling and enhances the electron pockets at the Q points.  Simultaneously, we observe a nearly-commensurate \tbt\ modulation that matches a predicted \tbt\ EPC-driven CDW due to phonon softening at the M point.  As has been a feature of several previous reports of CDW phases in TMDs~\cite{Ugeda2016,Rossnagel2011}, it is not possible to fully describe all of our data with \emph{either} a FSN \emph{or} EPC model independently.

We suggest that the coexistence of the \tbt\ and \trt\ phases is possible in part due to the effect of band folding.  We illustrate the reciprocal space vectors of the \trt\ periodicity ($u_1$ and $u_2$) and \tbt\ periodicity ($v_1$ and $v_2$) in Fig.~\ref{fig:brill}b, and it can be seen that these are simple linear combinations of one another (e.g., $v_1=u_1+u_2$).  Thus, band folding established by the shorter periodicity (illustrated by the hexagonal tiling in Fig.~\ref{fig:brill}b) will result in the band minimum at Q and the soft phonon mode at M being mapped back onto the zone center, providing an opportunity for the two mechanisms to couple. The fact that the two modulations are simple linear combinations of one another also provides an explanation for why the Fermi level gap we measure in STS spectra contains \tbt\ as well as \trt\ periodicity.

We have presented the discovery of a CDW ground state in \kmos, which is also the first observation of a CDW phase in a $d^2$ (column 6) TMD. Owing to the unique band structure of \kmos\ compared to $d^0$ and $d^1$ TMDs, our results provide new insight into the formation mechanisms of CDWs.  We observe a metal-insulator transition at 85~K, a $\sim25$~meV energy gap centred at the Fermi level, and \tbt\ and \trt\ periodicities that can be explained, respectively, by EPC and FSN mechanisms.  The \trt\ periodicity is observed exclusively near defects, suggesting that the coexistence of FSN and EPC phases can be attributed to their delicate sensitivity to the band edge and the strength of the interlayer electronic coupling.  Moreover, our observations suggest that CDW phases might be discoverable in other column 6 TMDs such as MoSe$_2$, WS$_2$, and WSe$_2$~\cite{Gusakova2017} at high electron-doping, and whose investigation might provide further insight into the physics of CDW formation. 

\begin{acknowledgement}
This research was financially supported by Engineering \& Physical Sciences Research Council (EP/L002140/1).  
\end{acknowledgement}
\begin{suppinfo}
The Supporting Information is available free of charge on the ACS Publications website at DOI: XXXX. STM images of \trt\ modulations at defect sites acquired at 5.5, 10, and 77~K.  STM images and analysis of phase slip boundaries.  Spatially-resolved tunnelling spectroscopy of \trt\ modulation. The data created during this research are openly available via zenodo.org at https://doi.org/10.5281/zenodo.3696856.
\end{suppinfo}

%%%%%%% Bibtex BIBLIOGRAPHY %%%%%%%%
%\bibliography{/Users/steven/academic/tex/bib/library}

\providecommand{\latin}[1]{#1}
\makeatletter
\providecommand{\doi}
  {\begingroup\let\do\@makeother\dospecials
  \catcode`\{=1 \catcode`\}=2 \doi@aux}
\providecommand{\doi@aux}[1]{\endgroup\texttt{#1}}
\makeatother
\providecommand*\mcitethebibliography{\thebibliography}
\csname @ifundefined\endcsname{endmcitethebibliography}
  {\let\endmcitethebibliography\endthebibliography}{}

\end{document}